# Lattice Experiments using Fermionic Operators and the Variational Eigensolver in a Quantum Computer

Wladimir Silva, Dept. of Computer Engineering, North Carolina State University, Raleigh NC.

This work describes a series of experiments in IBM's 16-qubit Guadalupe quantum processor to find the ground state of various lattice systems implemented in the Qiskit library. We aim to design a Variational Quantum Eigensolver (QVE) resistant to noise and independent of the number of vertices in the lattice. Furthermore, we test our solution against two Ising models very important in the study of critical points and phase transitions of magnetic systems as well as high-temperature superconductors, and quantum magnetism and charge density. We provide complete result metrics including final energies, precision percentages, execution times, angular parameters and source code for experimentation. Let's get started.

## I.     Foundation: The Ising Models

We seek to find ground states of critical points and phase transitions for common statistical mechanical models for which the spin or angular momentum is treated quantum mechanically. Our experiment focuses on the Heisenberg and Fermi-Hubbard models, which are very helpful in solid-state physics for describing transitions between conducting and insulating systems.

### A. Heisenberg Spin Model

Qiskit Nature, the package used in the experiment, implements the Heisenberg model for the study of critical points and phase transitions of magnetic systems [1]. It is given by the Hamiltonian:

$$H = -J \sum_{i,j} X_i Y_i Z_i \otimes X_j Y_j Z_j - h \sum_i X_i Y_i Z_i$$

where $X = \begin{bmatrix} 0 & 1 \\ 1 & 0 \end{bmatrix}, Y = \begin{bmatrix} 0 & -i \\ i & 0 \end{bmatrix}, Z = \begin{bmatrix} 1 & 1 \\ 0 & -1 \end{bmatrix}$ are the Pauli spin-1/2 matrices, J is the coupling constant (the strength exerted in an interaction), and *h* is the external magnetic field. The spectrum of this Hamiltonian describes the statistical properties of a system in thermodynamic equilibrium.

### B. Fermi Hubbard Model

Fermi Hubbard's describes electron interactions in solids, and is commonly used in the study of high-temperature superconductors as well as magnetism and charge density at the atomic scale [2]. In this model, there are two competing forces at play: a kinetic force, which pushes an atom to tunnel or hop to neighboring atoms, and a potential force which pushes it away from its neighbors. Therefore, its Hamiltonian has two terms:

$$H = \sum_{i,j} \sum_{\sigma=\uparrow\downarrow} t_{i,j}\, c^{\dagger}_{i,\sigma} c_{i,\sigma} + U \sum_i n_{i,\uparrow} n_{i,\downarrow}$$

where $t_{i,j}$ is the Hermitian interaction matrix, $c^{\dagger}_{i,\sigma}, c_{i,\sigma}$ are the creation, and annihilation operators of the fermion for *i* particles and spin σ, and U is the strength of the on-site interaction.

## II. The Lattices

The Ising models provide the foundation, next we seek an experiment to find the ground state of a subset of the lattice system provided by Qiskit Nature, more specifically:

**Table 1 Lattice samples used for experimental results.**

| | |
|---|---|
| 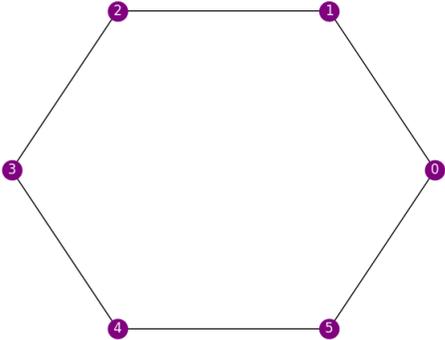 | Linear lattices: A linear set made of six and 16 qubits with a periodic boundary condition. This type is commonly used in computer simulations and mathematical models. Examples of this structure include crystalline minerals such as Carbonate Dolomite, Quartz and Beryl. [3] |
| 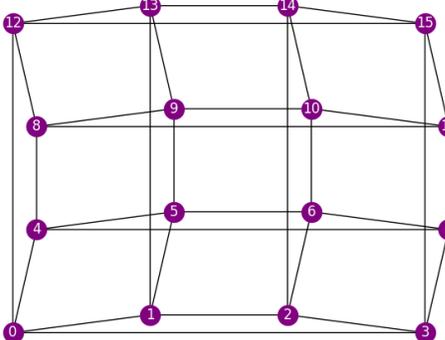 | Square Lattices: A 16 qubit square lattice with periodic boundaries that can be seen in metallic compounds such as iron, chromium, tungsten, aluminum, copper, gold and silver as well as crystalline solids such as carbon, silicon, and germanium. [4] |

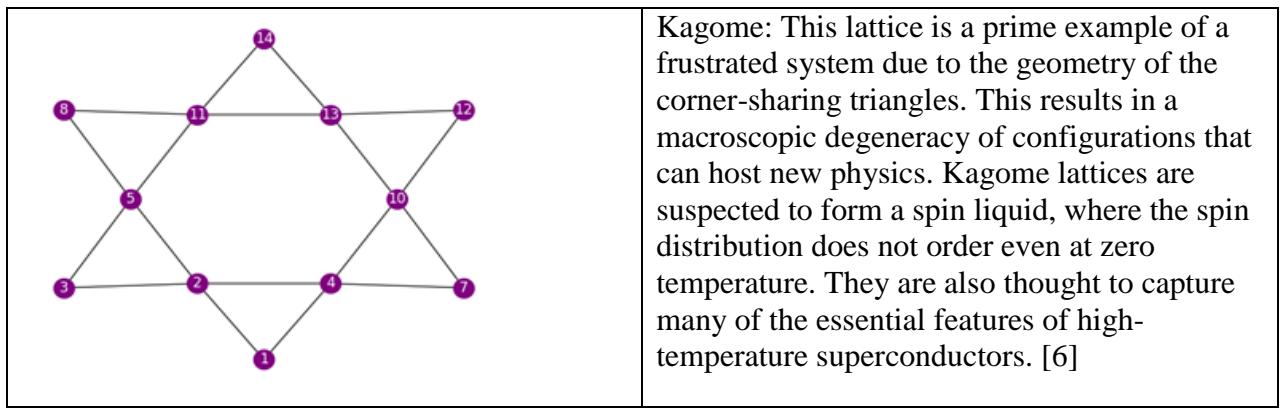

Kagome: This lattice is a prime example of a frustrated system due to the geometry of the corner-sharing triangles. This results in a macroscopic degeneracy of configurations that can host new physics. Kagome lattices are suspected to form a spin liquid, where the spin distribution does not order even at zero temperature. They are also thought to capture many of the essential features of high-temperature superconductors. [6]

## III. A Target Guided VQE

The Variational Quantum Eigensolver is the de-facto algorithm to find ground states in a quantum computer. In the chemistry domain this process is known as state preparation which requires a set of common components including:

- Ansatz: This is the initial quantum circuit at the heart of the VQE. Our experiment uses EfficientSU2 (special unitary group of degree 2) from Qiskit circuit library. This Ansatz uses a heuristic pattern that can be used to prepare trial wave functions common in machine learning. In our initial tests, EfficientSU2 achieved the highest efficiency compared with others due to its simplicity (see figure 1). It is important to keep in mind that Control-X gates have errors around 1% and these errors accrue, thus a circuit with an excessive number of gates will be detrimental to the final result.

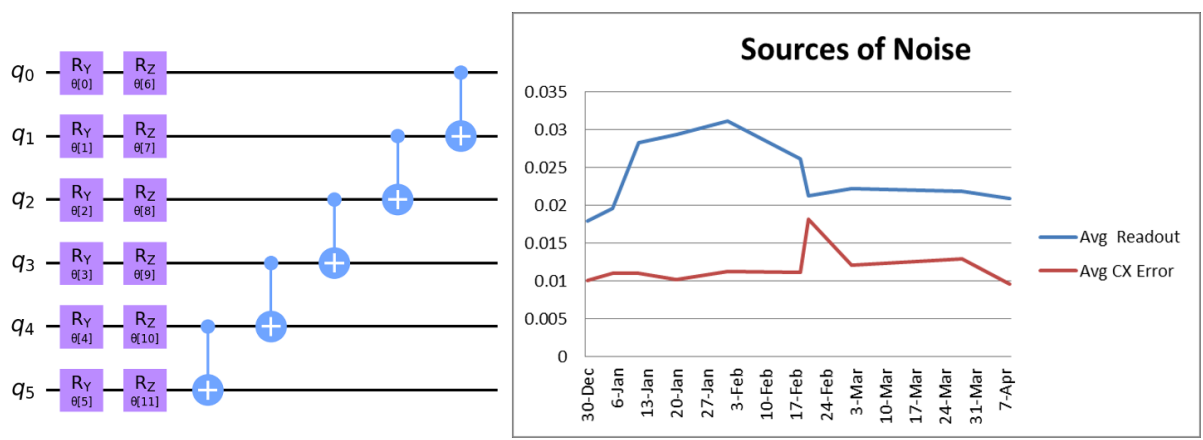

**Figure 1** EfficientSU2 Ansatz for 6 qubits from Qiskit circuit library (left). On the right, the two main sources of noise from the Guadalupe quantum processor taken over a period of 4 months during this experiment. Readout (measurement) noise accrues at a rate close to 2%, almost double the rate of a Control-X gate.

- Optimizer: The optimizer is the classical component of the VQE. Its job is to feed the Hamiltonian to the Ansatz to estimate the expected energy value. Qiskit implements an extensive library of optimizers. In our early tests, the Nakanishi-Fujii-Todo (NFT) algorithm performed the best due to its sequential minimal optimization method for quantum-classical hybrid algorithms, which converges faster, and is robust against statistical error. [5]

We use a customized implementation of the VQE that zig-zags around the expected target energy by correcting the uniform interaction of the Hamiltonian using recursive techniques. See listing 1 for a description of the pseudo code.

| Input | Initialization |
| --- | --- |
| lattice | The problem lattice. |
| expected_energy | Optional desired ground state, calculated by classical means. This value is used to guide the VQE towards the ground state. If empty, the dynamic Hamiltonian correction will not apply. |
| model | The Ising model: Heisenberg or FermiHubbard. |
| ham | The problem Hamiltonian, |
| UI | Uniform interaction initialized to the weight of the edge of the lattice. |
| threshold | Desired error threshold (1% by default). |
| optimizer | VQE optimizer: NFT by default. |

```
function compute_minimum_eigenvalue(ansatz, lattice) {
    // Objective function
    function objective(thetas) {
        if task_done
            return last_value

        // Execute quantum circuit
        job = estimator.run(ansatz, ham, thetas)

        if fix_hamiltonian {
            ham           = get_hamiltonian(model, lattice, weight)
            fix_hamiltonian = False
        }
        // Get results from the quantum job
        est_result = job.result()

        // Get the measured energy value
        last_value = value = est_result.values
```

```
        // if below the target (expected_energy), add a bias (0.25) to the Hamiltonian UI
        if expected_energy != None {
            if ( value < expected_energy ) {
                delta = abs(value) - abs(expected_energy)
                if ( delta > 10 ) {
                    if abs(self._weight) > 1.0 {
                        weight        -= 0.25
                        fix_hamiltonian = True
                    }
                }
            }
            // If we reach the error threshold, return collected result
            if expected_energy != 0 {
                err    = 100 * compute_relative_error(self._expected_energy, value)
                if err <= threshold
                    task_done = True
            }
        }
        // Send result back to the client using callback function
        if callback is not None:
            callback(value)

        return value
    }

    // Select an initial point (thetas) for the ansatzs' parameters
    if initial_point is None
        x0 = pi/4 * random(ansatz.num_parameters)
    else
        x0 = initial_point

    // Run optimization
    result   = optimizer.minimize(objective, x0=x0)

    if expected_energy != None {
        delta    = abs(result.value) - abs(expected_energy)
        err      = 100 * compute_relative_error(expected_energy, result.value)

        // If ended above the target, recurse w/ a higher weight (bias)
```

```
        if err > threshold and recursions < 5 and (not task_done) {
            recursions   += 1   // track recursive calls
            if delta < 0
                weight   += abs(delta)/100 + 0.3
            else
                weight   -= abs(delta)/100 + 0.3

            ham          = get_hamiltonian(model, lattice, weight)
            initial_point = result.initial_point

            // recursive call
            compute_minimum_eigenvalue(ham, lattice)
        }
    }
    // Populate VQE result & return
    return result
}
```
**Listing 1: Pseudo code for the target guided implementation of the VQE.**

## IV. Transpilation makes a big difference

Transpilation is an optimization technique used by IBM's quantum processors to match the topology of a specific device, resulting in circuit rewriting to fit hardware constraints and optimizations for performance. This is a complicated process that can achieve big savings in execution time and results precision (see figure 2).

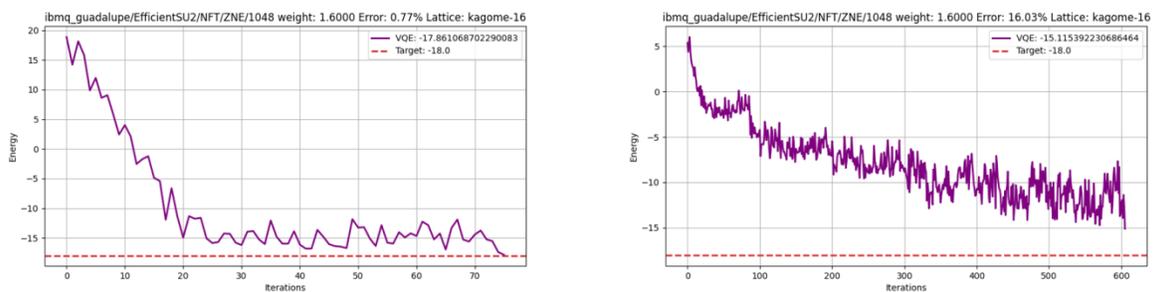

**Figure 2** Results from the ground state solution for the Kagome lattice on IBM's 16 qubit Guadalupe processor: On the left, result from a transpiled Ansatz with high precision and relatively low number of cycles. On the right, the un-transpiled result showing significantly lower precision with high number of cycles, a difference of 500+.

Transpilation resulted in huge cost savings in our case: consider table 2 with the final metrics from two runs of the VQE to find the ground state of the Kagome from the experiment in figure 2.

Table 2 Kagome lattice ground state results for a transpiled vs. un-transpiled Ansatz: Transpilation run for over 1h with a high precision (below 1%). Un-transpiled resulted in an 11h run with a disappointing 16% relative error.

| Ansatz: EfficientSU2 (reps=1, entanglement='linear') | Ansatz: EfficientSU2 (reps=1, entanglement='linear') |
|---|---|
| Optimizer : NFT(maxiter=100) | Optimizer : NFT(maxiter=100) |
| Resilience : Zero Noise Extrapolation (ZNE) | Resilience : ZNE |
| Expected ground state energy: -18.0 | Expected ground state energy: -18.0 |
| Computed ground state energy: -17.86106870 | Computed ground state energy: -15.11539223 |
| **Relative error: 0.77%** | **Relative error: 16.025%** |
| **Execution Time: 1.41h** | **Execution Time: 11.34h** |

## V.     Experimental Results

All experimental results run on hardware (16-qubit Guadalupe processor) using the EfficientSU2 Ansatz, NFT optimizer and edge weight of 1.6 with Zero Noise Extrapolation for quantum resilience.

### Heisenberg Model Results for 12-Qubits

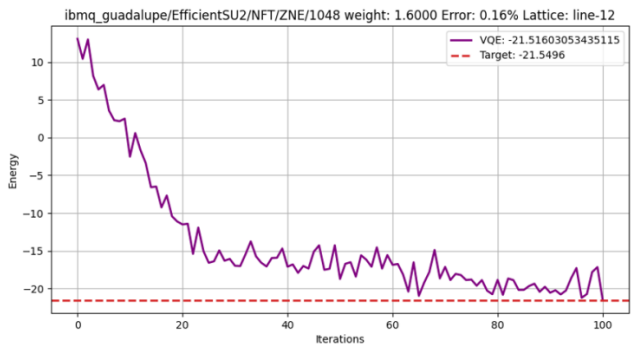 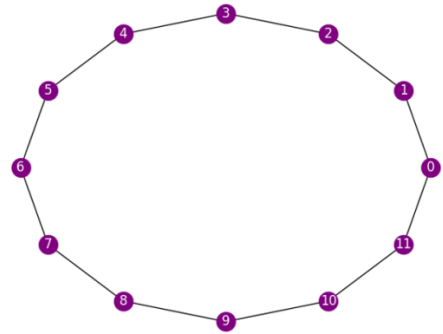

- Lattice: linear with periodic bounds
- Computed ground state energy: -21.51603053
- Expected ground state energy: -21.54960000
- Relative error: 0.15577767%
- Execution time: 0.67h

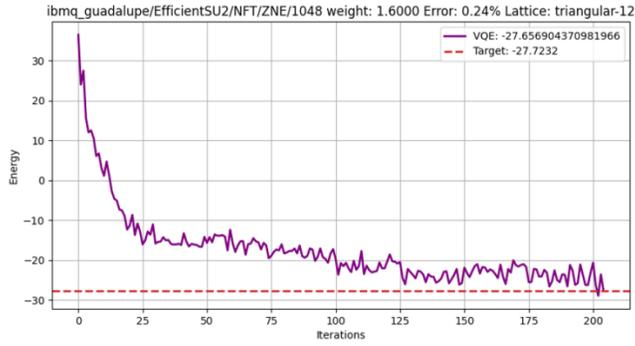 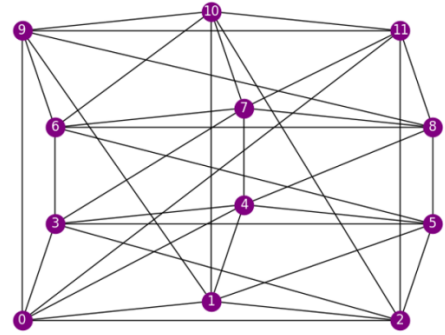

- Lattice: triangular with periodic bounds
- Computed ground state energy: -27.65690437
- Expected ground state energy: -27.72320000
- Relative error: 0.23913412%
- Execution time: 2.43h

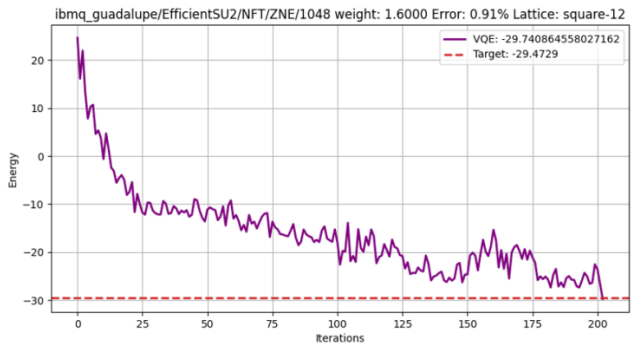 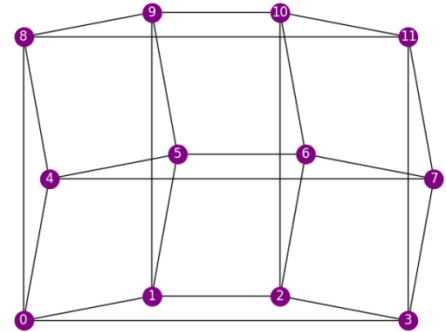

- Lattice: square with periodic bounds
- Computed ground state energy: -29.74086456
- Expected ground state energy: -29.47290000
- Relative error: 0.90918966%
- Execution time: 1.43h

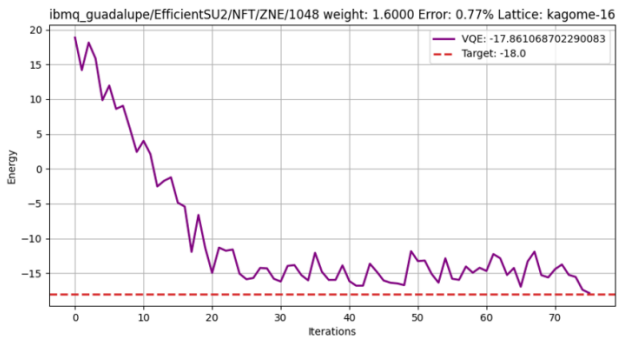 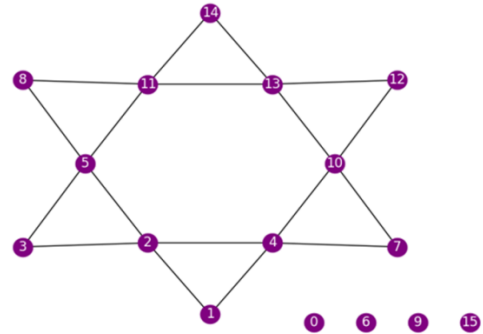

- Lattice: Kagome (transpiled)
- Computed ground state energy: -17.86106870
- Expected ground state energy: -18.00000000
- Relative error: 0.77184054%
- Execution time: 1.41h

**Heisenberg Model Results for 16-Qubits**

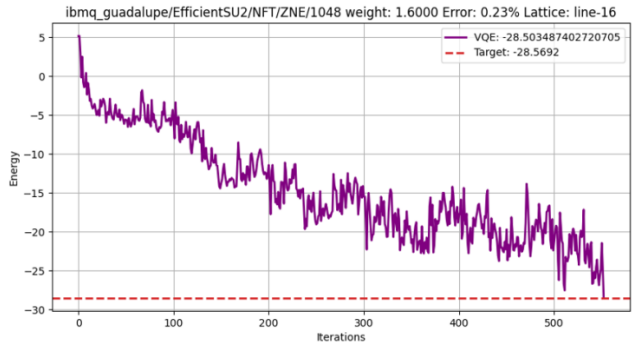
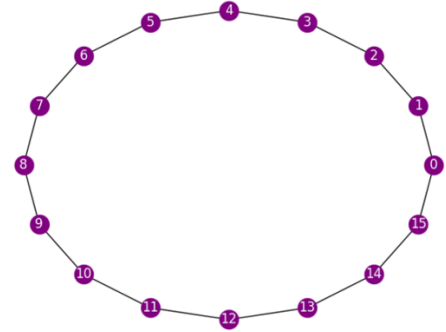

- Lattice: linear with periodic bounds
- Computed ground state energy: -28.50348740
- Expected ground state energy: -28.56920000
- Relative error: 0.23001203%
- Execution time: 2.66h

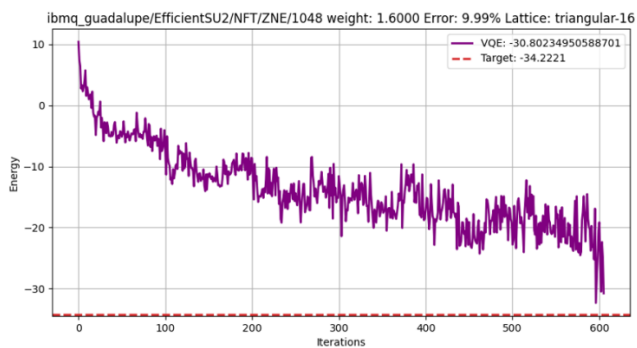
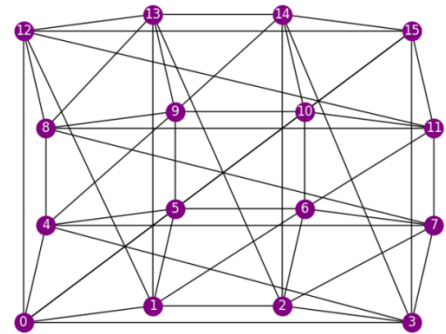

- Lattice: triangular with periodic bounds
- Computed ground state energy: -30.80234951
- Expected ground state energy: -34.22210000
- Relative error: 9.99281311%
- Execution time: 8.32h

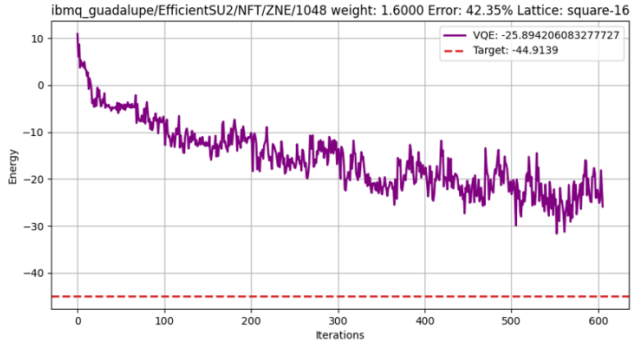
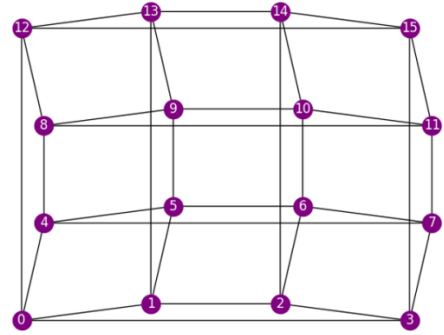

- Lattice: square with periodic bounds
- Computed ground state energy: -25.89420608
- Expected ground state energy: -44.91390000
- Relative error: 42.34701043%
- Execution time: 5.42h

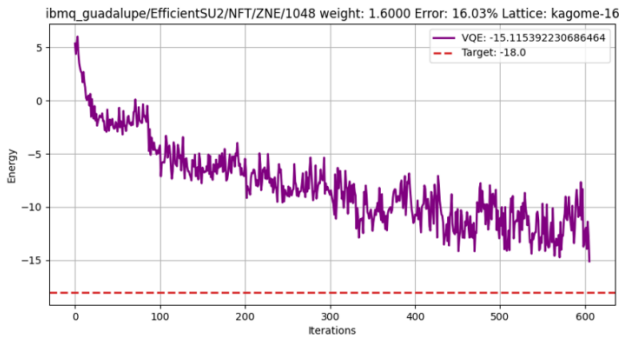
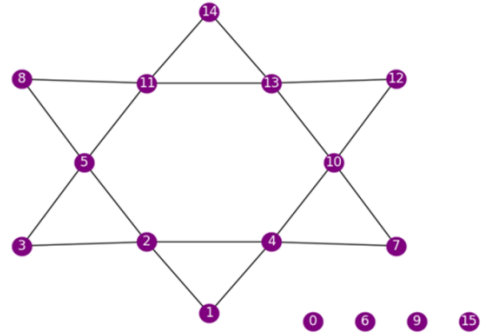

- Lattice: Kagome (un-transpiled)
- Computed ground state energy: -15.11539223
- Expected ground state energy: -18.00000000
- Relative error: 16.02559872%
- Execution time: 11.34h

**Fermi-Hubbard Results for 6-qubits**

For the Fermi-Hubbard stage of the experiment, we use the Jordan-Wigner mapper to transform the fermionic operators to the qubit space. Jordan-Wigner maps the occupation of one spin-orbital to the occupation of one qubit. Thus from Fermions to Qubits the creation-annihilation operators become:

$$\{a_i, a_i\} = 0, \{a_i^\dagger, a_i^\dagger\} = 0, \{a_i, a_i^\dagger\} = \delta_{i,j} \ (Ferminions)$$

$$[\sigma_i, \sigma_i] = 0, [\sigma_i^\dagger, \sigma_i^\dagger] = 0, [\sigma_i, \sigma_i^\dagger] = \delta_{i,j} (Qubits)$$

A side effect of Jordan-Wigner is that the resulting Hamiltonian uses twice as many qubits as vertices in the lattice. Thus for 12 and 16 vertex lattices we would need 24 and 32 qubits respectively. Therefore we run our experiment using 6 vertex lattices due to the 16-qubit constraint of the Guadalupe processor.

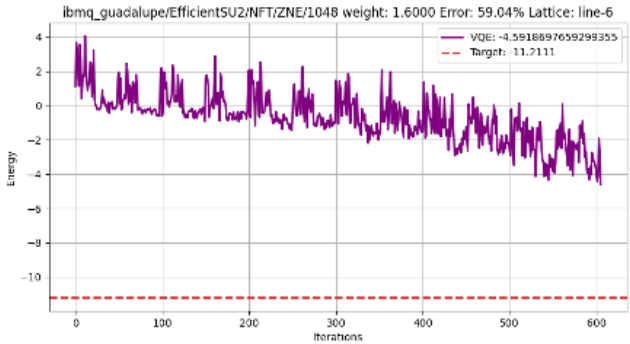 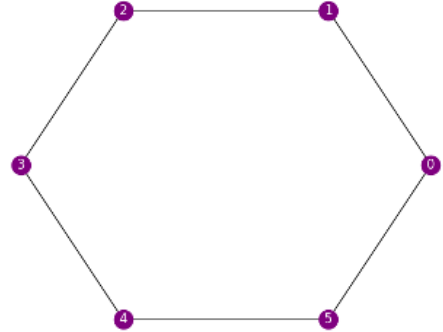

- Lattice: linear with periodic bounds
- Computed ground state energy: -4.59186977
- Expected ground state energy: -11.21110000
- Relative error: 59.04175535%
- Execution time: 3.56h

Note on the experiments for the triangular, an square lattices: The experiments failed after multiples tries with messages such as "Internal Server Error", "Failed - queue error: session is closed". At the end, we were not able to collect any results on these.

## VI. Conclusion

Our customized VQE performed well, achieving high precisions and low run times for relatively low number of qubits on multiple lattice types; however as the number of qubits increased above 12, precision degraded significantly and run times increased exponentially. Due to the qubit constraints of the quantum processor we are not able to go above 16 qubits. Nevertheless, there is room for improvement and the results are encouraging enough to continue the quest for a fault tolerant, general purpose VQE that can model large number of elements. This type of VQE is of critical importance because no classical eigensolver is capable of modeling large numbers of elements due to the tremendous requirements of memory and computational power (see table 3).

**Table 3 Run times for the classical eigensolver in Qiskit's python library on a quad core Pentium processor with 32GB of RAM.**

| Number of Qubits | Run time (s) | Memory Consumption (MB) |
|---|---|---|
| 12 | 2 | 4 |
| 13 | 12 | 8 |
| 14 | 116 | 9,384 |
| 15 | 1233 | 16,384 |

| 16 | Crash | Crash |

## VII. Acknowledgment

We would like to thank IBM and North Carolina State University's Quantum Hub for the time and access provided to the quantum processor to run these experiments. You are welcome to tryout the source code for this manuscript online at https://github.com/Shark-y/quantum_lattice/